\documentstyle[12pt,psfig]{article}
\advance\textheight by\topskip

\def\degr{\hbox{$^\circ$}}

\def\arcsec{\hbox{$^{\prime\prime}$}}

\begin{document}

\title{ Radio-optical identification of very-steep spectrum radio sources
              from the UTR-2 catalogue
}

\author{
Heinz Andernach
\and {\it Depto.\,de Astronom\'\i a, Univ.\ de Guanajuato, Guanajuato, Mexico}
\and Oleg V. Verkhodanov
\and {\it Special Astrophysical Observatory, Nizhnij Arkhyz, Russia}
\and Natalia V. Verkhodanova
\and {\it Special Astrophysical Observatory, Nizhnij Arkhyz, Russia}
}

\date{}
\maketitle

\begin{abstract}
\renewcommand{\baselinestretch}{1.2}
We used radio source catalogues accessible from the CATS database
({\tt\small http://cats.sao.ru}) 
to establish radio continuum spectra for decametric radio sources
in the UTR-2 catalogue (Braude et al.\ 1978--1994). From these,
we select a sample of 23 sources with ultra steep radio spectra
($\alpha\le-$1.2, S\,$\sim\nu^{\alpha}$) and present accurate 
positions and sizes from FIRST and NVSS.
The search for optical counterparts from the APM (object) and DSS (image)
databases, as well as infrared and X--ray identifications of these 
UTR sources are in progress.
\end{abstract}

\renewcommand{\baselinestretch}{1.0}

\section {Introduction}

A radio survey obtained with the UTR telescope (Kharkov, Ukraine) 
at frequencies 10--25\,MHz has resulted in a catalog of 1822 sources 
(Braude et al.\ 1978--1994; {\tt www.ira.kharkov.ua/UTR2}).
Covering about 30\% of the sky north of $-$13\degr\ declination, 
this survey is presently the lowest-frequency source catalog 
of its size, and thus provides an ideal basis to study the little known
optical identification content of sources selected at decametric 
frequencies.
In the original version of the UTR-2 catalog (UTR in what follows)
there is no radio identification at other frequencies for 7\% of the sources,
and for 81\% there is no optical identification. Our goal is to 
identify all UTR sources with known radio sources and to search for
optical counterparts on the Digitzed Sky Surveys.

Here we present our first results on a subsample of
ultra-steep spectrum (USS) sources (spectral index $\alpha\le-$1.2, 
 ~S\,$\sim\nu^{\alpha}$). This class of sources is being actively studied by 
various groups (Parijskij et al.\ 1996; R\"ottgering et al.\ 1997; 
McCarthy et al.\ 1997), mainly because they are often identified with
very distant radio galaxies, which are probes of the early Universe
and thought to be indicators of proto-clusters (e.g.\ Djorgovski 1987).

\section {Radio identification}

The rather large uncertainties of UTR positions ($\sim$0.7\degr) require
an iterative process for finding radio counterparts at successively
higher frequencies (and thus higher positional accuracy). In this we aided 
ourselves by selecting previously cataloged sources from the 
CATS database (Verkhodanov et al.\ 1997a) in a box of
RA$\times$DEC~=~40$'\times$40$'$ centred on the nominal UTR position.
The ``raw'' spectra given by these fluxes were refined using
computer charts of source locations around UTR positions.
All counterparts from TXS, GB6 and PMN within circles of 1$'$ radius were
considered one source.  Groups of sources lying further apart
were assigned separate spectra, each with the UTR flux as upper limit.

We were able to fit spectra for all but 7 of the 2314 radio counterparts to
UTR sources. Fits were either straight (S), convex (C$^{-}$),
or concave (C$^{+}$) curves in the lg\,$\nu$--lg\,S plot
(see also Verkhodanov et al.\ 1997b, 1998).
The resulting catalog includes information from a large number of
electronically available catalogs of radio, infrared,
optical and X--ray sources.
The distribution of radio source spectra among the various spectral types
is given in Table~1, and Fig.~1 shows the spectral index distribution
of sources at high and low Galactic latitudes.

\begin{table}[h!]
\caption{Distribution of radio continuum spectral types of 
2307 radio counterparts to UTR sources, where X=log$_{10}$(frequency/MHz), 
and Y=log$_{10}$(flux density in Jy)}
\begin{center}
\begin{tabular}{|llrr|}
\hline
Spectral class & Fitting function & N & \% \\
\hline
Straight (S)        & $Y = +A+B*X                   $ &  894 &  39 \\
Convex   ($C^{+}$)  & $Y = +A{\pm}B*X-C*X^2         $ &  184 &   8 \\
Concave  ($C^{-}$)  & $Y = +A-B*X+C*X^2             $ & 1150 &  50 \\
~~~~~~~~~~~~~~~~or  & $Y = {\pm}A{\pm}B*X+C*EXP(-X) $ &   79 &   3 \\
\hline
\end{tabular}
\end{center}
\end{table}

\begin{figure}[h!]
\centerline{
\hbox{
\psfig{figure=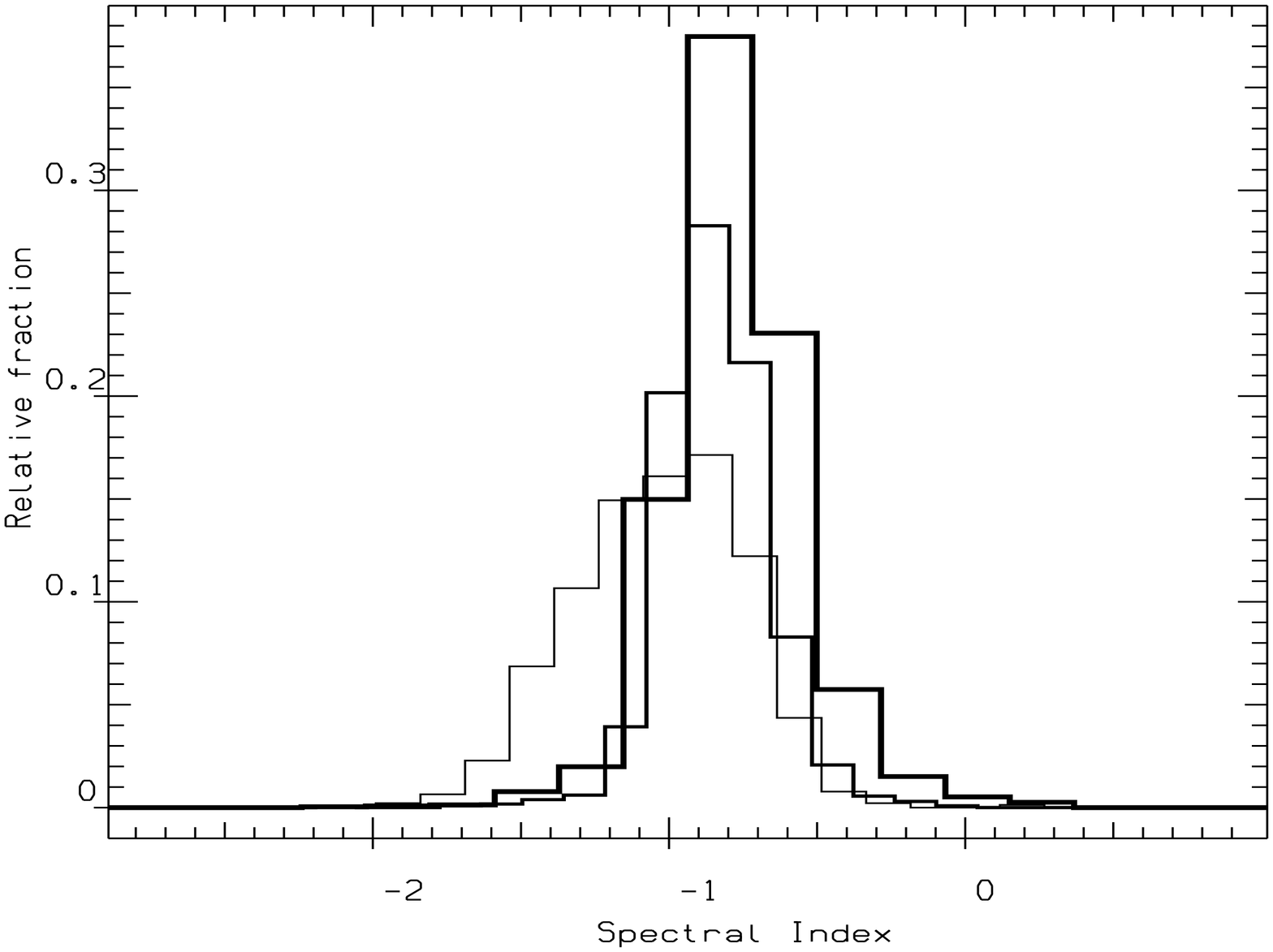,width=10cm,height=7.4cm,angle=-90}
\psfig{figure=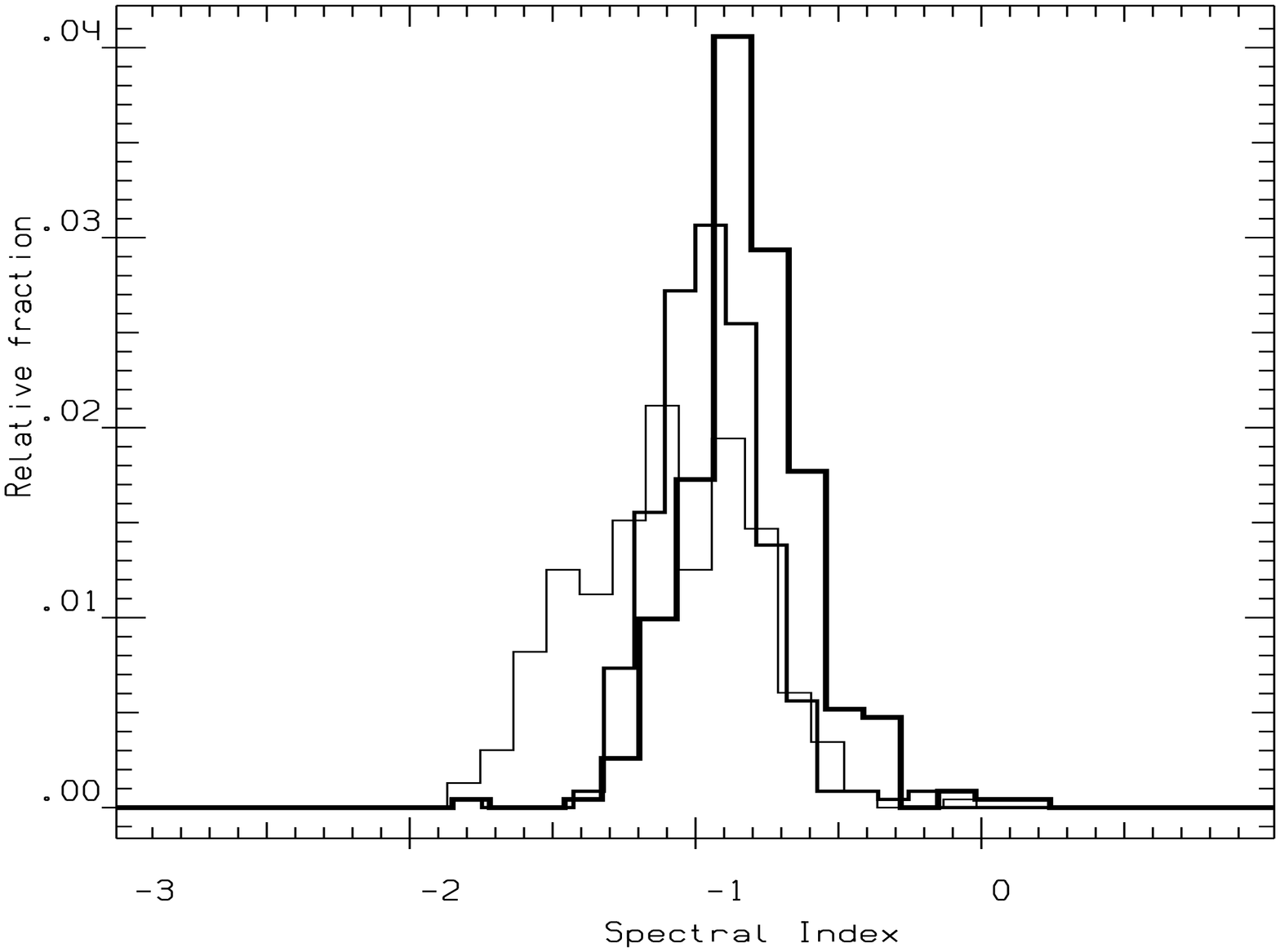,width=10cm,height=7.4cm,angle=-90}
}}
\caption{
Distributions of spectral indices of 2307 radio counterparts to
   2314 UTR sources, fitted
at frequencies 80 (thin line), 365 and 1400 MHz (bold line).
Left: 2004 high Galactic latitude
sources ($|b|>10^\circ$); right: 303 sources at low Galactic latitudes
($|b|<10^\circ$).}
\end{figure}
{\small
\begin{table}
\begin{center}
\caption{ The 38 FIRST counterparts to 23 USS sources ($\alpha\le-$1.2) 
from the UTR catalog. We list the UTR name, the size of the source complex
(or deconvolved size if a single component) from NVSS and FIRST, respectively,
the FIRST RA \& Dec, the peak and integrated component flux at 1.4\,GHz from FIRST, 
their deconvolved major and minor axes, and major axis position
angle (N through E) from FIRST.}
\begin{tabular}{|lrrcrrrrr|}
\hline
 UTR name     & NVSS~ & FIRST & RA ~(J2000)~ Dec  &  $S_p$~~ &  $S_i$~~ & Maj   &  Min &  PA\\
 ~~~(B1950)     & size ('') &  size ('') &  &  (mJy) & (mJy) & (\arcsec)~ &(\arcsec)~ & (\degr) \\
\hline
GR0002+00~b    & 16.  &  7.4 &000650.56$+$003648.4 &   47.6 &   75.9&   7.4 &   1.1 &  154 \\  
GR0135$-$08    & 28.  & 20.~~&013714.87$-$091155.4 &    5.2 &   44.3&  30.8 &   7.5 &   98 \\  
               &      &      &013715.08$-$091203.3 &   10.2 &   36.9&  15.6 &   4.3 &   90 \\  %
               &      &      &013715.45$-$091155.8 &    8.2 &   25.7&  13.0 &   5.4 &  148 \\  %
               &      &      &013716.22$-$091149.5 &   10.0 &   39.3&  11.3 &   9.3 &  119 \\  %
GR0257$-$08~a  &$<$18.&  4.5 &025919.15$-$074501.2 &  162.5 &  211.9&   4.5 &   1.3 &   59 \\  
GR0257$-$08~b  & 80.  & 83.~~&030040.22$-$075302.2 &   18.4 &   52.8&  15.6 &   3.0 &  174 \\  
               &      &      &030040.56$-$075259.6 &   10.1 &  122.8&  30.7 &  13.1 &  163 \\  %
	       &      &      &030042.99$-$075413.8 &    7.0 &  34.2 &  16.2 &   8.4 &   32 \\
	       &      &      &030042.99$-$075358.0 &   13.6 &  47.8 &  12.3 &   7.4 &  152 \\
	       &      &      &030043.60$-$075418.3 &    6.4 &  22.0 &  11.2 &   7.6 &   75 \\
	       &      &      &030043.75$-$075407.2 &    6.6 &  22.4 &  15.6 &   4.8 &  170 \\
GR0723+48~b    & 47.  &  3.1 &072651.18$+$474041.5 &   23.8 &   29.1&   3.1 &   2.0 &  175 \\  
	       &      &      &072655.00$+$474051.0 &   51.8 &   75.7&   5.6 &   1.1 &   85 \\
GR0818+18      & 24.  & 22.~~&082032.48$+$192731.3 &   31.1 &   49.2&   6.1 &   1.7 &  174 \\  
               &      &      &082032.73$+$192709.0 &   76.4 &   98.5&   3.9 &   1.6 &    8 \\  %
GR0858$-$02~b  & 70.  & 55.~~&085935.06$-$015842.1 &   11.7 &   27.2&  10.2 &   3.2 &  122 \\  
               &      &      &085936.10$-$015851.8 &    9.8 &   15.9&   6.2 &   2.8 &  128 \\  %
               &      &      &085938.21$-$015908.1 &   20.2 &   43.6&   8.1 &   4.5 &  133 \\  %
GR0910+48      & 36.  & 29.~~&091359.00$+$482738.0 &   26.9 &   88.6&  11.6 &   5.2 &  100 \\  
               &      &      &091401.83$+$482729.2 &   39.6 &  110.2&  10.9 &   3.9 &  106 \\  %
GR0922+42~b    &$<$19.&  2.7 &092559.66$+$420335.3 &  199.7 &  244.1&   2.7 &   2.4 &   98 \\  
GR0942+54      & 14.  &  8.~~&094618.12$+$543003.8 &   51.1 &   57.1&   2.5 &   0.9 &   36 \\  
               &      &      &094618.53$+$543010.1 &   75.3 &   80.6&   1.7 &   1.1 &   16 \\  %
GR1149+42      &$<$17.&  5.0 &115213.58$+$415344.9 &   83.7 &  115.4&   5.0 &   1.0 &   18 \\  
GR1214$-$03    &$<$17.&  3.2 &121755.30$-$033722.0 &  176.9 &  208.9&   3.2 &   1.2 &  111 \\  
GR1223$-$00    & 25.  & 25.~~&122722.97$-$000813.8 &    5.8 &   13.6&   8.4 &   5.0 &  100 \\  
               &      &      &122724.54$-$000821.1 &    8.4 &   16.9&   7.1 &   4.6 &  116 \\  %
GR1243+04      &$<$18.&  6.8 &124538.38$+$032320.1 &  249.4 &  379.9&   6.8 &   1.3 &  158 \\  
GR1318+54      & 25.  & 23.~~&132202.81$+$545758.1 &   30.3 &   90.3&   9.5 &   5.9 &   82 \\  
               &      &      &132205.33$+$545805.3 &    8.6 &   35.9&  10.7 &   8.6 &   55 \\  %
GR1320+43      &$<$19.&  3.3 &132232.32$+$425726.5 &  129.3 &  154.6&   3.3 &   1.0 &   81 \\  
GR1355+01~c    &$<$19.& 10.~~&135821.64$+$011442.0 &   77.0 &   88.4&   2.9 &   1.4 &   42 \\  
               &      &      &135822.08$+$011449.4 &  169.9 &  180.9&   1.6 &   1.3 &   32 \\  %
GR1447+57      &$<$19.&  4.5 &144630.04$+$565146.8 &   75.2 &   99.2&   4.5 &   0.7 &  148 \\  
GR1527+51~b    &$<$19.&  1.4 &152828.36$+$513401.4 &  203.9 &  212.8&   1.4 &   0.8 &  140 \\  
GR1539+53~b    &$<$18.&  6.4 &154144.69$+$525054.5 &   87.0 &  136.7&   6.4 &   0.8 &   66 \\  
GR1613+49      &$<$19.&  3.0 &161631.16$+$491908.2 &   56.1 &   66.2&   3.0 &   1.3 &   10 \\  
GR1731+43      & 42.  & 40.~~&173333.87$+$434318.6 &   32.9 &   57.9&   6.3 &   3.0 &  164 \\  
               &      &      &173334.26$+$434300.8 &   22.8 &   27.3&   3.1 &   1.5 &  173 \\  %
               &      &      &173334.41$+$434251.4 &   12.3 &   15.8&   3.3 &   2.4 &  159 \\  %
               &      &      &173334.58$+$434239.8 &   27.3 &   51.0&   7.5 &   2.3 &  150 \\  %
GR2211$-$08~b  &$<$19.&  8.6 &221519.65$-$090005.8 &   75.2 &  125.0&   8.6 &   1.4 &  176 \\  
\hline         
\end{tabular}
\end{center}
\end{table}
}

\section {Subsample of ultra-steep spectrum sources}

In our catalog of 2314 radio counterparts
(the full list will be published in a forthcoming paper)
there are 422 S-type sources with ``very-steep spectrum'' (VSS), and
for the present work we selected from these a subsample of 102 
``ultra-steep spectrum'' (USS) objects ($\alpha\le-$1.2).
To further increase the radio-positional accuracy, we searched for
radio counterparts of USS sources in the February (1998) version of
the FIRST catalog
(White et al.\ 1997), resulting in 38 FIRST counterparts for 23 UTR sources 
(see Table~2).
If a UTR source has more than one acceptable counterpart in FIRST,
we label these components with letters a, b, c, etc.  
Only one of the FIRST components (labeled GR1527+51\,b) is 
truly unresolved by the FIRST beam of $\sim$\,5$''$ 
(i.e.\ has a major and minor axis of $<$\,2$''$), while all other 
objects have a multi-component or extended structure.
We checked the sources also in the lower resolution 
NVSS at 1.4\,GHz (Condon et al.\ 1998). Usually, the larger
the source complex, the larger the NVSS/FIRST flux ratio.
Radio spectra of some of the source complexes are shown in Fig.~2.
Examples of two FIRST maps of multi-component objects 
GR0910+48 and GR0942+54 overlaid on DSS-2 images are shown in Fig.~3.
According to the NASA Extragalactic Database (NED), the 
complex source GR0135$-$08
is identified with the $z=0.041$ galaxy MCG-02-05-020,
GR1214$-$03 is an LCRS QSO at z=0.184, and  
GR1243+04 is a radio galaxy (4C+03.24) at z=3.57.
Our investigation of DSS-2 images for the unidentified
sources is in progress.

\begin{figure}
\centerline{
\psfig{figure=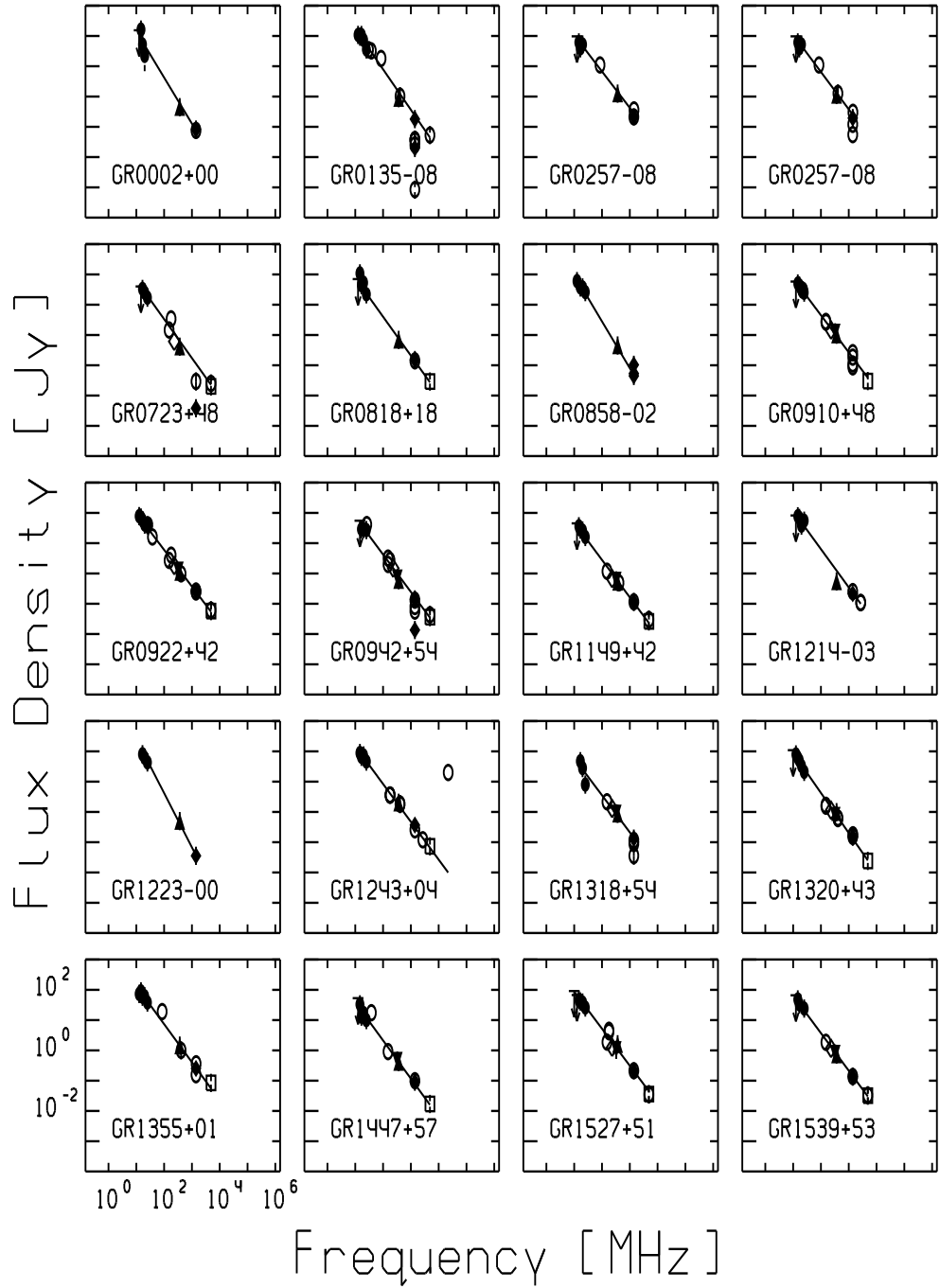,width=16cm}
}
\caption{Radio spectra of 20 USS sources from UTR, identified with FIRST sources.}
\end{figure}

\vspace*{3mm}

\begin{figure}
\centerline{
\vbox{
\psfig{figure=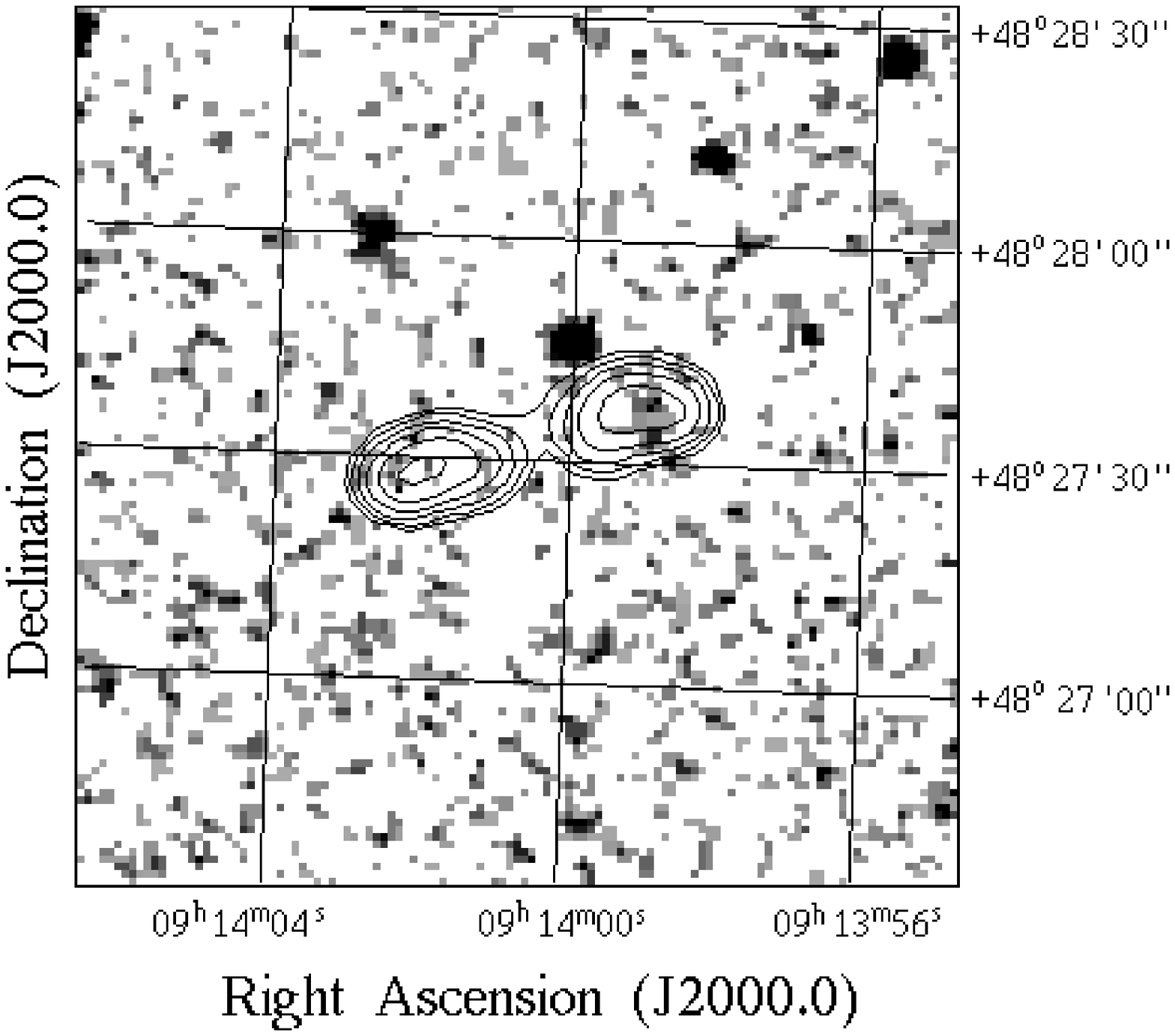,width=10cm}
\psfig{figure=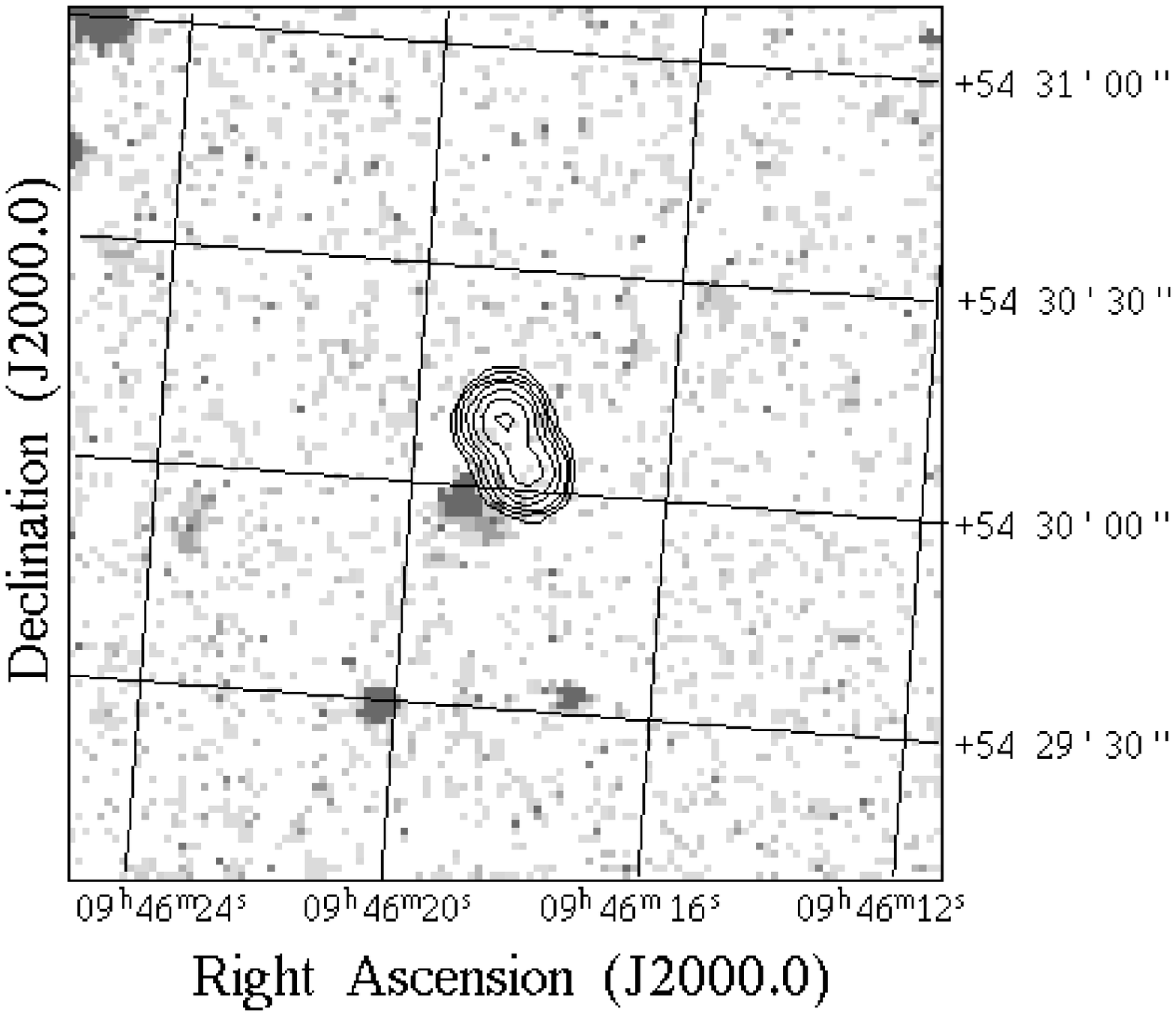,width=10cm}
}}
\caption{
Examples of two FIRST maps of multicomponent objects GR0910+48 (upper panel)
and GR0942+54 (lower panel) overlaid on DSS-2 images.
There are no optical counterparts above the
plate limit at the symmetry center of these radio doubles.
}
\end{figure}

{\bf Acknowledgements}
We wish to thank A.P.\ Miroshnichenko and D.~Krivitskij
of Institute of Radio Astronomy (Kharkov) for providing data and for
useful discussion. 
The co-creators of the CATS database Sergei Trushkin and Vladimir Chernenkov 
(SAO RAS) provided useful comments.

{}

\end{document}